\documentclass[10pt]{article}

\usepackage[letterpaper]{geometry}
\usepackage{hicss}
\usepackage{times}

\usepackage[none]{hyphenat}
\usepackage{url}
\usepackage{latexsym}
\usepackage{minted}
\usepackage{indentfirst}
\usepackage{graphicx}
\graphicspath{{images/}}
\usepackage[
    style=apa,
  ]{biblatex}
\AtBeginBibliography{%
  \fontsize{9}{10.8}\selectfont
}

\usepackage{amsmath,amssymb,amsfonts}
\usepackage{booktabs}
\usepackage[caption=false,font=small]{subfig}
\usepackage[
    font={small,sf,bf},
    justification=centering,
    labelsep=period
]{caption}

\usepackage{threeparttable}
\usepackage{algorithmic}

\title{A Community Impact-Centered Network Reduction Method for Power Systems Resilience Planning}

\author{
  Jiarui Gao \\
  University of Michigan \\
  {\underline{jiaruii@umich.edu}}
  \And
  Abigail L. Beck \\
  University of Houston \\
  {\underline{albeck2@uh.edu}}
  \And
  Rabab Haider \\
  University of Michigan \\
  {\underline{rababh@umich.edu}}
}

\date{June 15, 2026}

\begin{document}

\maketitle
\begin{abstract}
Power systems resilience planning requires evaluating many hazard scenarios and investment portfolios, often requiring network reduction for computational tractability. The resulting aggregation, however, can obscure community impact data needed for equitable planning. We present a network reduction method that jointly preserves network physics and the spatial distribution of community data. We embed a household-level scarce resource score that characterizes outage impact within optimization-based Kron reductions with two formulations: a scarcity-weighted method that prioritizes vulnerable nodes, and a scarcity-constrained method that bounds data dissimilarity within aggregated clusters. Both are evaluated on a realistic distribution feeder of Galveston Island, Texas and benchmarked against multiple approaches. The scarcity-constrained method performs best, achieving 86\% reduction with worst-case voltage errors below 0.01~p.u. while preserving the spatial distribution of community data. By explicitly controlling data heterogeneity within and across clusters, the proposed method yields reduced networks that retain decision-relevant community information without sacrificing electrical accuracy.
\end{abstract}

\subsubsection*{Keywords:} 

optimal network reduction, Kron reduction, power systems resilience, community impact metrics, mixed integer linear programming

\section{Introduction}

Power systems are a critical infrastructure whose disruption during natural hazards such as extreme weather, flooding, or hurricanes, can lead to severe consequences for affected communities, for example, population dislocation and well-being loss. This is further exacerbated by the disproportionate impact grid outages have on disadvantaged communities \parencite{andresen2023understanding}. Resilience planning can mitigate the impact of power outages through pre-hazard strengthening, flexible operations, and post-hazard restoration. However, resilience is still an emerging concept in power systems planning with no agreed-upon framework for measuring it or guiding the investment of limited resource budgets under which grid planners must prioritize across communities \parencite{stankovic2022methods, zamuda2018energy}. Even when high-resolution community and grid data are available to grid planners, the resulting models must be solved for a large number of hazard scenarios and across various investment portfolios. Doing so on a high-fidelity network model of tens of thousands of nodes can be computationally challenging. To maintain computational tractability, reduced network models can be used. However, information loss from data aggregation and the use of generalized vulnerability indices can fail to protect certain high-risk communities, preventing effective resource allocation policies in power systems resilience planning \parencite{Pollack_2025_equitablewildfire}. In this paper, we begin to address this challenge. We propose a method for network reduction that maintains data integrity by preserving the spatial distribution of community impact metrics present in the original network, while maintaining computational tractability of the resulting resilience planning models. 

Recent studies have recognized the importance of incorporating community impact metrics into power resilience planning. \textcite{rodriguez2024power} develop a framework to mitigate overheating-risk inequity during heat waves. The community impact metric is first quantified at the individual-building level and then aggregated to the feeder-section level to represent the overall community risk. \textcite{toplu2024impact} develop a stochastic optimization planning model that incorporates inequities in well-being loss caused by flood-induced transmission outages. Community impact data at the census-tract level is mapped to each transmission bus. These works take reduced network representations as given a priori and use community data at varying levels of representation. However, reduced network representations derived from high-resolution datasets of network and community information are subject to an ecological fallacy, where population averages (i.e., at an aggregated cluster-level) are assumed to equally represent the individual members \parencite{piantadosi1988ecological, jargowsky2021disastrous}. The reduced network is also susceptible to the modifiable areal unit problem, a statistical bias common in spatial analysis wherein different groupings of distinct nodes can obscure local data extremes and artificially alter the visibility of underlying sociodemographic relationships \parencite{openshaw1984modifiable}. An additional challenge in power systems network reduction is maintaining the physical system behaviour (i.e., voltage-current relationship) of the network. In this work, we take household-level data and a high-resolution realistic grid model to produce a reduced network representation that jointly preserves the underlying power physics and any spatial distribution of community data present in the original network.

The Kron reduction is a standard tool used to obtain network-reduced models for power systems studies. This method eliminates a subset of interior network nodes using the Schur complement of the nodal admittance matrix, while maintaining system behaviour (e.g., voltage–current relationships) of the remaining nodes \parencite{dorfler2013kron}. While able to preserve the physics of the network, it is limited in its reduction capability in practical distribution systems and requires eliminated nodes to be pre-specified. \textcite{chevalier2022optikron} propose an optimization-based Kron reduction to iteratively identify nodes for removal while minimizing the resulting voltage error. We extend this framework to account for the data integrity of community impact metrics.

We propose a community impact-centered network reduction method for power systems resilience planning and evaluate it on a realistic distribution feeder of Galveston Island, Texas. The contributions of this paper are as follows. First, we develop a power systems model for the distribution network on Galveston Island, Texas. We extend an existing utility pole infrastructure dataset, producing a network that can be used for both structural resilience analysis and power systems studies. Second, we propose a community impact-centered optimization-based network reduction method to reduce the network while preserving electrical accuracy and the spatial distribution of community impact data. We adopt the metric of scarce resource scores to characterize the potential impact of a power outage on individual households, and formulate two methods that integrate this metric into the objective function (a \emph{scarcity-weighted} method) and constraints (a \emph{scarcity-constrained} method). We define two evaluation metrics: the \emph{within-cluster scarcity variance} to measure the similarity of scores within grouped nodes, and the \emph{between-cluster scarcity variance} to measure the spatial distribution across the network. Third, we validate the proposed method against three existing network reduction methods: Kron reduction, relaxed Kron reduction, and Opti-KRON. The results demonstrate that the proposed approach can carry out network reduction while preserving both electrical accuracy and the spatial distribution of community data present in the original network.

The remainder of this paper is organized as follows. Section 2 presents the network reduction method. Section 3 describes the Galveston Island case study and the evaluation metrics. Section 4 presents the numerical results and discusses the main findings, and conclusions are provided in Section 5.

\section{Methodology}
\subsection{Electrical Network Representation}
Consider a distribution network as a graph $\mathcal{G}(\mathcal{N}, \mathcal{M})$, with $\mathcal{N}$ the set of $n=|\mathcal{N}|$ nodes and $\mathcal{M}$ the set of $m=|\mathcal{M}|$ edges. The network topology and grid physics are described as
\begin{equation}
    I = YV ,
    \label{eq:nodal}
\end{equation}
where $Y\in\mathbb{C}^{n\times n}$ is the admittance matrix, and $V\in\mathbb{C}^{n}$ and $I\in\mathbb{C}^{n}$ denote the nodal voltage and current injection vectors in rectangular coordinates. When the admittance matrix is nonsingular, the corresponding impedance matrix is defined as $Z = Y^{-1}$.

\subsection{Kron Reduction}
Let $\mathcal{N}$ be partitioned into the set of kept nodes $\mathcal{K}$ and eliminated nodes $\mathcal{R}$. Reordering the nodal variables according to this partition, \eqref{eq:nodal} can be written as

\begin{equation}
\begin{bmatrix}
I_k\\
I_r
\end{bmatrix}
=
\begin{bmatrix}
Y_{kk} & Y_{kr}\\
Y_{rk} & Y_{rr}
\end{bmatrix}
\begin{bmatrix}
V_k\\
V_r
\end{bmatrix},
\label{eq:kron_partition}
\end{equation}
where the subscripts $k$ and $r$ denote kept and removed nodes, respectively. Eliminating $V_r$ from \eqref{eq:kron_partition} gives
\begin{equation}
    I_k =
    \left(
    Y_{kk}-Y_{kr}Y_{rr}^{-1}Y_{rk}
    \right)V_k
    +Y_{kr}Y_{rr}^{-1}I_r .
    \label{eq:kron_general}
\end{equation}
The Kron reduction restricts eliminated nodes to zero-injection nodes. With $I_r=0$ the reduced system representation is
\begin{align}
    I_k &= Y_{\mathrm{red}}V_k ,
    \label{eq:kron_reduced} \\
    Y_{\mathrm{red}} =&
    Y_{kk}-Y_{kr}Y_{rr}^{-1}Y_{rk}.
    \label{eq:yred}
\end{align}
The network representation in \eqref{eq:kron_reduced} results in an equivalent network that maintains the current-voltage relationship of the kept nodes, however the set of eliminated nodes is restricted to those with zero current injection. To further reduce the network size, the elimination criteria can be relaxed, to permit the removal of selected nonzero-injection nodes by neglecting their respective current injections. The  resulting reduced model introduces current and voltage mismatches relative to the original network, quantified by
\begin{align}
    \Delta I_k &= Y_{kr}Y_{rr}^{-1}I_r
    \label{eq:current_mismatch} \\
    \Delta V_k &=
    Y_{\mathrm{red}}^{-1}\Delta I_k .
    \label{eq:voltage_mismatch}
\end{align}
The \textit{relaxed Kron reduction} approximates the exact Kron reduction, but requires sets $\mathcal{K}$ and $\mathcal{R}$ to be specified in advance.

\subsection{Optimization-based Kron Reduction}
Optimization-based Kron reduction methods generalize this approach by formulating an optimization problem to identify node sets $\mathcal{K}$ and $\mathcal{R}$. \textcite{chevalier2022optikron} propose Opti-KRON, a mixed-integer linear program that iteratively reduces the network size while constraining introduced voltage deviations. Opti-KRON offers modeling flexibility by optimizing the trade-off between network reduction and electrical accuracy using tunable weighting parameters to capture user-defined criteria. However, this formulation focuses solely on electrical accuracy. We extend the Opti-KRON method to jointly preserve electrical accuracy and the spatial distribution in community impact metrics across nodes.

\subsection{Metric Definition: Scarce Resource Scores} 
We use scarce resource scores, $R$, to characterize the potential impact of a power outage on individual households. This metric captures characteristics that have been shown to predispose households to potential increased impact during a power outage.  Households with higher potential impact during an outage hinder community resilience gains in preparing for, responding to, or recovering from the outage \parencite{beck2026equity}. For each household $h$, the scarce resource score is computed from four binary household indicators:
\begin{equation}
R_h = 5 - \left(B_h + D_h + M_h + L_h\right),
\label{eq:household_resource}
\end{equation}
where $B_h$, $D_h$, $M_h$, and $L_h$ indicate renter status, presence of dependents, minority status, and low-income status, respectively. These indicators were selected for the Galveston Island case study after characterizing the household sociodemographic data (see Section 3.1). Each indicator takes a value of one if the specific condition is met, and zero otherwise. A lower value of $R_h$ signifies that household $h$ is likely more vulnerable to an outage and adverse outage impacts. 

Households are assigned to infrastructure nodes in $\mathcal{G}$ according to geographic proximity. Let $\mathcal{H}_i$ denote the set of households assigned to node $i$. The node-level scarce resource score $R_i$ is defined as the average of the household scores assigned to the node, as
\begin{equation}
    R_i =
    \frac{1}{|\mathcal{H}_i|}
    \sum_{h\in\mathcal{H}_i} R_h .
    \label{eq:node_resource}
\end{equation}
After network reduction, the cluster-level scarce resource score is the average of the node-level scores assigned to the cluster. Let $\mathcal{C}_j$ denote the set of nodes assigned to cluster $j$. The scarce resource score of cluster $j$ is calculated as
\begin{equation}
 \bar{R_j}=
\frac{1}{|\mathcal{C}_j|}
\sum_{i\in\mathcal{C}_j} R_i.
\label{eq:cluster_resource}
\end{equation}

\subsection{Scarcity-Weighted Network Reduction}
The scarcity-weighted network reduction iteratively solves an optimization problem to identify node sets $\mathcal{K}$ and $\mathcal{R}$ to reduce the network size while preserving network voltages and the spatial distribution of scarce resource scores. 

At iteration $t$ of the network reduction method, let $\mathcal{G}^{(t)}$ denote the graph of the input network, with $\mathcal{N}^{(t)}$ denoting the set of $n^{(t)}=|\mathcal{N}^{(t)}|$ nodes. The scarcity-weighted network reduction at iteration $t$ determines the assignment matrix $A^{(t)}$ to produce a reduced network $\mathcal{G}^{(t+1)}$, subject to a maximum voltage error $\delta^{(t)}$. The binary assignment matrix is defined as $A^{(t)}\in\{0,1\}^{n^{(t)}\times n^{(t)}}$, where $A^{(t)}_{ij}=1$ if node $j\in\mathcal{N}^{(t)}$ is eliminated and assigned to representative node $i\in\mathcal{N}^{(t)}$, and $A^{(t)}_{ij}=0$ otherwise. The diagonal entry $A^{(t)}_{ii}$ indicates whether node $i$ is retained in the reduced network representation as a representative node for the node cluster ($A^{(t)}_{ii}=1$) or is eliminated ($A^{(t)}_{ii}=0$).  The node sets $\mathcal{K}^{(t)}$ and $\mathcal{R}^{(t)}$ are recovered from $A^{(t)}$ where $\mathcal{K}^{(t)} = \{i \in \mathcal{N}^{(t)} | A_{ii}^{(t)} = 1 \}$ and $\mathcal{R}^{(t)} = \mathcal{N}^{(t)} \setminus \mathcal{K}^{(t)}$. For notational simplicity, the superscript $(t)$ is omitted in the following formulation. 

\begin{subequations}
\label{eq:social_object}
\setlength{\jot}{0pt}
\begin{align}
\min_{A,\delta}\quad
    & \delta - \frac{\alpha}{n}
    \sum_{j=1}^{n} \bar{R}_j(1-A_{jj})
    \label{eq:social_obj_only}\\
\mathrm{s.t.}\quad
    & \sum_{i\in\mathcal{N}} A_{ij}=1,
    \; \forall j\in\mathcal{N},
    \label{eq:opt_assign}\\
    & A_{ij}\leq A_{ii},
    \; \forall i,j\in\mathcal{N},
    \label{eq:opt_rep}\\
    & \widetilde{V}=ZAI,
    \label{eq:representative_voltage}\\
    & \delta \leq \gamma,
    \label{eq:opt_gamma}\\
    & \widetilde{V}_{i}^{\mathrm{re}}-V_{j}^{\mathrm{re}}
    \leq \delta + M_V(1-A_{ij}),
    \; \forall i,j\in\mathcal{N},
    \label{eq:opt_vre_upper}\\
    & V_{j}^{\mathrm{re}}-\widetilde{V}_{i}^{\mathrm{re}}
    \leq \delta + M_V(1-A_{ij}),
    \; \forall i,j\in\mathcal{N},
    \label{eq:opt_vre_lower}\\
    & \widetilde{V}_{i}^{\mathrm{im}}-V_{j}^{\mathrm{im}}
    \leq \delta + M_V(1-A_{ij}),
    \; \forall i,j\in\mathcal{N},
    \label{eq:opt_vim_upper}\\
    & V_{j}^{\mathrm{im}}-\widetilde{V}_{i}^{\mathrm{im}}
    \leq \delta + M_V(1-A_{ij}),
    \; \forall i,j\in\mathcal{N},
    \label{eq:opt_vim_lower}\\
    & \sum_{j\in\mathcal{N}} (1-A_{jj}) \leq \beta n,
    \label{eq:opt_budget}\\
    & A_{ij}\leq E_{ij},
    \; \forall i,j\in\mathcal{N},
    \label{eq:opt_conn}\\
    & A_{ij}\in\{0,1\},\quad \delta\geq0,
    \; \forall i,j\in\mathcal{N}.
    \label{eq:opt_binary}
\end{align}
\end{subequations}

The objective function in \eqref{eq:social_obj_only} balances the voltage error $\delta$ against a scarcity-weighted network reduction reward. The reduction incentive $\alpha/n$ is scaled by the cluster-level scarce resource score $\bar{R}_j$. This formulation penalizes the removal of resource-constrained nodes (indicated by lower $R$ scores) to preserve the households' potential vulnerability to an outage. Lower scores indicate higher vulnerability, so their removal yields smaller optimization rewards.

Constraints \eqref{eq:opt_assign} and \eqref{eq:opt_rep} require that every reduced node $j$ be assigned to exactly one retained node $i$. Constraint \eqref{eq:representative_voltage} calculates the voltage of the representative nodes $\widetilde{V}$ by mapping the system impedance matrix $Z$ to the aggregated current injections $AI$. Constraints \eqref{eq:opt_vre_upper}--\eqref{eq:opt_vim_lower} bound the real and imaginary voltage error of nodes assigned to each cluster by $\delta$. The binary assignment is relaxed using a big-M formulation with parameter $M_V$ \parencite[see][Lemma 2]{chevalier2022optikron}. Constraint \eqref{eq:opt_gamma} restricts the allowable voltage error resulting from network reduction. To maintain computational tractability, \eqref{eq:opt_budget} limits the fraction of nodes removed at each iteration using the parameter $\beta \in [0,1]$. Constraint \eqref{eq:opt_conn} preserves geographic proximity of nodes by restricting the assignment of node $j$ to adjacent nodes $i$ using the connectivity matrix $E$, where $E_{ij}=1$ indicates that nodes $i$ and $j$ are directly connected. 

In the first iteration $t=1$, the scarce resource scores are initialized as $\bar{R}_j^{(t)} = R_j$, where $R_j$ denotes the original node-level scarce resource score. For subsequent iterations $t > 1$, the active node set $\mathcal{N}^{(t)}$ is updated to include only the retained nodes, and the network matrices ($Z$, $I$, and $E$) are dimensionally reconstructed. This iterative reduction process continues until no further nodes are eliminated, i.e., $\mathcal{R}^{(t)}=\emptyset$.
\subsection{Scarcity-Constrained Network Reduction}

While the \textit{scarcity-weighted network reduction} modifies the retention preference of nodes, it does not explicitly control the heterogeneity of scarce resource scores among nodes assigned to a cluster. We propose an improved formulation, the \textit{scarcity-constrained network reduction}, which introduces a constraint to restrict the $R$-dissimilarity between an eliminated node $j$ and the representative node $i$ of its assigned cluster, to a user-defined tolerance $\phi$. The network reduction problem then jointly optimizes over the scale of network reduction, and the voltage errors and $R$-dissimilarity in each cluster of the reduced network. At iteration $t$ of the network reduction method, let $\mathcal{G}^{(t)}$ denote the graph of the input network, and $A^{(t)}$ denote the assignment matrix. For notational simplicity, the superscript $(t)$ is once again omitted in the following formulation. 

\begin{subequations}
\label{eq:joint_social_model}
\setlength{\jot}{1pt}
\begin{align}
\min_{A,\delta,\eta}\quad
    & \delta + \omega \eta
    - \frac{\alpha}{n}\sum_{j\in\mathcal{N}} \bar{R}_j(1-A_{jj})
    \label{eq:joint_social_obj}\\
\mathrm{s.t.}\quad
    & \eqref{eq:opt_assign}-\eqref{eq:opt_binary},
    \label{eq:joint_existing_constraints} \\
    & D_{ij}
    =
    \max_{p\in\mathcal{C}_i,\;q\in\mathcal{C}_j}
    |R_p-R_q|,\notag\\
    &\qquad \forall \{ i, j\in\mathcal{N} \mid E_{ij}=1, i\neq j\},
    \label{eq:joint_Dij_def}\\
    & \eta \leq \phi,
    \label{eq:joint_eta_bound}\\
    & D_{ij}
    \leq
    \eta + M_R\left(1-A_{ij}\right),\notag\\
    &\qquad \forall i,j\in\mathcal{N},\ i\neq j .
    \label{eq:joint_social_bigM}
\end{align}
\end{subequations}

The objective function in \eqref{eq:joint_social_obj} introduces a $R$-dissimilarity penalty term $\omega\eta$, where $\eta$ represents the maximum $R$-dissimilarity across the clusters and parameter $\omega$ weights the preservation of spatial distribution in community information against voltage accuracy. Constraint \eqref{eq:joint_Dij_def} defines $D_{ij}$ as the maximum pairwise $R$-dissimilarity between a node in the $i$-th cluster ($p\in C_i$) and a node in the $j$-th cluster ($q\in C_j$) of the network $\mathcal{G}^{(t)}$. Given the input network $\mathcal{G}^{(t)}$, the values of $D_{ij}$ can be precomputed prior to solving the optimization problem. Constraint \eqref{eq:joint_eta_bound} restricts the allowable $R$-dissimilarity  resulting from network reduction. Constraint \eqref{eq:joint_social_bigM} bounds the $R$-dissimilarity of nodes assigned to each cluster by $\eta$. The binary assignment is relaxed using a big-M formulation with parameter $M_R$. In the first iteration $t=1$, the maximum pairwise $R$-dissimilarity $D_{ij}$ is the difference in scarce resource scores between neighbouring nodes. 

\section{Experimental Setup}

\subsection{Test Network: Galveston Island, Texas}
\begin{figure}[]
    \centering
 {%
        \includegraphics[width=\columnwidth]{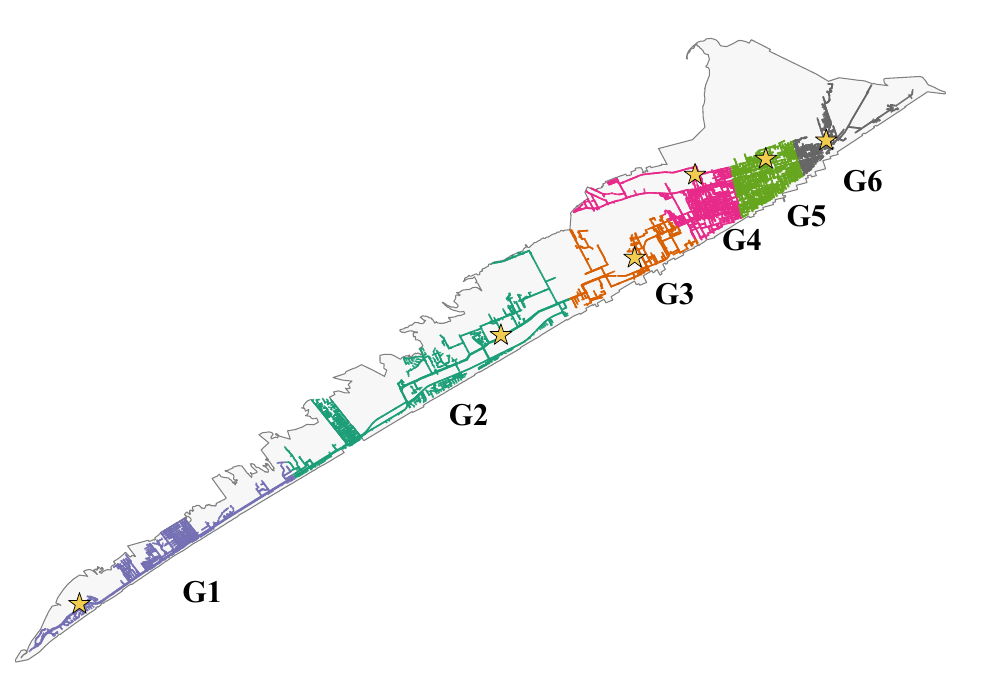}
    }
    \caption{Distribution network topology for Galveston Island, Texas. The six substations for feeders G1 to G6 are annotated with a star icon.}
    \label{fig:all_feeders}
\end{figure}

The proposed community impact-centered network reduction methods are evaluated on a realistic distribution grid model of Galveston Island, Texas \parencite{darestani2022galveston}. The dataset describes the location and specifications of distribution poles, distribution network connectivity, and locations of substation connecting the distribution and transmission grids. We extend the pole-level structural dataset to a an electrical grid model, providing a single network upon which both structural resilience analysis and electrical impact assessment can be conducted. The Galveston Island network is divided into six regions, with each distribution network serviced by a distribution substation. We evaluate our methodology on Zone G3, a feeder with 1,457 poles and residential loads. Figure~\ref{fig:all_feeders} shows the network.

\paragraph{Electrical network model} We develop an electrical network model of the distribution grid, where each pole represents a node, and each line segment connecting two poles is modeled as an edge in the graph $\mathcal{G}$. The distribution substation is treated as the slack bus and cannot be eliminated during network reduction. The branch parameters are adapted from the IEEE 33-bus distribution test case, assuming a balanced three-phase network, and scaled to the G3 network so the nodal voltages satisfy the standard voltage range of $0.95$--$1.05$ p.u. The resulting resistance and reactance parameters are $r = 0.03~\Omega/\mathrm{km}$ and $x = 0.035~\Omega/\mathrm{km}$.

\paragraph{Household sociodemographic and load data} We use the housing unit allocation \parencite{rosenheim2021integration} from the IN-CORE Galveston testbed to retrieve high-resolution household sociodemographic data at the level of housing units within single and multi-family residential structures \parencite{incore_galveston}. The binary household indicators for Eqn. \eqref{eq:household_resource} are defined using the thresholds from \cite{beck2026equity}, as described next. The presence of dependents status $D_h$ is met when a household has more than two dependents. The minority status $M_h$ is met for non-White households or white household that are Hispanic. The low-income status is met for households with annual income below the mean income of the G3 region. Each housing unit is assigned to the nearest pole according to geographic proximity. Building load data are obtained for Galveston County from NLR's U.S. residential load profiles \parencite{openei_load}, by matching building type to the structure ID in the Galveston dataset. A load snapshot at 2018-04-01 14:00:00 is selected to represent a typical operating condition. Nodal loads are computed by aggregating the household loads assigned to each pole. The resulting network has $n=1,457$ of which 600 are nonzero-injection nodes. On average, each node (representing a pole in the network) is associated with 11.74 housing units, with average nodal load of 10.8 kW, and a maximum nodal load of 296 kW.

\paragraph{Power flow and baseline voltages} We use \texttt{PowerModelsDistribution.jl} \parencite{pmd} to solve power flow. The per-unit (p.u.) base values are set to $ S_{\mathrm{base}} = 1.0~\mathrm{MVA}, V_{\mathrm{base}} = 12.66~\mathrm{kV}$, 
where $V_{\mathrm{base}}$ is the line-to-line voltage base. The corresponding single-phase phase-to-neutral voltage base is $ V_{\mathrm{base,ph}} = \frac{12.66}{\sqrt{3}} = 7.309~\mathrm{kV}$. The maximum nodal voltage is $0.994$ p.u., and the minimum nodal voltage is $0.966$ p.u. The resulting network voltage and current profiles are used as the reference values for implementing and evaluating the network reduction methods.

\subsection{Implementation of Network Reduction Methods}
The Kron reduction is applied as a baseline method on the Galveston G3 network, denoted as $\mathcal{G}$. It removes zero-injection nodes while retaining the slack node and all nonzero-injection nodes, producing a reduced network $\mathcal{G}_{Kron}$. Since the scarce resource score is defined only for nodes with residential load, the two networks $\mathcal{G}$ and $\mathcal{G}_{Kron}$ are equivalent, fully capturing the current-voltage relationship of the network and spatial distribution of community impact metrics.

The relaxed Kron reduction is implemented as a structural approach to network reduction. The G3 network is classified into three topological layers: primary, secondary, and branch layers. Starting from the substation, the longest continuous radial path is identified as the primary layer. Nodes connected to the primary layer that serve as branching points, together with the downstream paths extending from them to the next terminal or branching point, are classified as the secondary layer. The remaining downstream nodes are classified as branch-layer nodes. Nodes directly connected to the primary layer without downstream branches are also assigned to the branch layer. We evaluate two relaxed Kron reduction cases. In the first case (Relaxed Kron-I) zero-injection nodes are removed and both primary- and secondary-layer nodes are retained. In the second case (Relaxed Kron-II)
zero-injection nodes are removed and only primary-layer nodes are retained. These two cases provide different reduction levels and allow the effect of retaining additional topological detail to be examined.

The reduced network $\mathcal{G}_{Kron}$ is used to initialize the three iterative optimization-based methods (Opti-KRON, scarcity-weighted, and scarcity-constrained). At each iteration, representative nodes are retained, eliminated nodes are merged into their corresponding representative clusters, and the network matrices, current injections, and connectivity matrix are updated. The optimization model is then solved again on the updated reduced network until the stopping criterion is reached.

\subsection{Parameter Selection for Optimization-based Methods}
The Opti-KRON and scarcity-weighted methods are solved with $\alpha=0.001$, network reduction of 25\% (i.e., $\beta=0.25$), and maximum voltage error $\gamma=0.01$ p.u. These values were selected from parameter tuning, to balance the number of iterations, retained network size, and to ensure the voltage error $\delta$ is tightly bounded by the prescribed tolerance $\gamma$. The scarcity-constrained method is solved with $\alpha=0.05$, network reduction of 25\% (i.e., $\beta=0.25$), maximum voltage error $\gamma=0.01$ p.u., maximum $R$-dissimilarity limit $\phi=2.0$, and error weighting parameter $\omega=0.005$. These values provide a suitable tradeoff between the scale of network reduction, and the voltage errors and $R$ dissimilarity in each cluster of the reduced network. 

\subsection{Evaluation Metrics}
The following metrics are used to evaluate the reduction methods.

\paragraph{Network Reduction Rate}
This metric quantifies the degree of network reduction, as
\begin{equation}
    \rho =
    \frac{n_0-n_{\mathrm{red}}}{n_0}\times 100\%,
    \label{eq:reduction_rate}
\end{equation}
where $n_0$ is the number of nodes in the original network $\mathcal{G}^{(1)}$ and $n_{\mathrm{red}}$ is the number of kept nodes in the reduced network, $\mathcal{G}^{(T+1)}$ after $T$ iterations. A larger value of $\rho$ indicates a higher reduction level.

\paragraph{Voltage Error}
This metric quantifies the electrical approximation error introduced by network reduction. For the relaxed Kron reduction, the voltage error is evaluated as the voltage mismatch of the kept nodes $\mathcal{K}$ relative to the original network as in Eqn. \eqref{eq:voltage_mismatch}. For the optimization-based methods where eliminated nodes $j\in\mathcal{R}$ are clustered, and each cluster has a representative node $i\in\mathcal{K}$, the voltage error compares the voltage of each eliminated node in $\mathcal{G}^{(1)}$ with its cluster voltage, as
\begin{equation}
    e_V^{\max} =
    \max_{i,j:\, A_{ij}=1}
    \left|
    |V_j| - |\widetilde{V}_{i}|
    \right|,
    \label{eq:max_voltage_deviation}
\end{equation}
where $i$ denotes the representative node of a cluster, $j$ denotes an eliminated node assigned to representative node $i$, $V_j$ is the original voltage of node $j$, and $\widetilde{V}_{i}$ is the voltage estimate for the cluster represented by $i$. The voltage error is reported in \emph{milli}-per unit, or \emph{m}p.u. Smaller values of $e_V^{\max}$ indicate lower voltage errors and better preservation of the current-voltage relationship.

\paragraph{Spatial Distribution of Scarce Resource Score}
We define two metrics to evaluate the impact of network reduction on the spatial distribution of the households' scarce resource scores: within-cluster $R$ variance and relative between-cluster $R$ variance. Consider the reduced network $\mathcal{G}^{(T+1)}$ after $T$ iterations, and denote $J=|\mathcal{N}^{(T+1)}|$ as the number of clusters. 

The \emph{within-cluster $R$ variance} measures the average variance of cluster-level scarce resource scores across the $J$ clusters. A smaller value indicates that nodes assigned to the same cluster have similar scarce resource scores. It is calculated as
\begin{equation}
    \sigma^2_{\mathrm{R,in}} =
    \frac{1}{J}
    \sum_{j=1}^{J}
    \left(
    \frac{1}{|C_j|}
    \sum_{i\in C_j}
    (R_i-\bar{R}_j)^2
    \right)
    \label{eq:within_social_variance}
\end{equation}

The \emph{between-cluster $R$ variance} measures the variance of the cluster-level scarce resource scores in the reduced network, defined as
\begin{equation}
    \sigma^2_{\mathrm{R,bw}} =
    \frac{1}{J}
    \sum_{j=1}^{J}
    \left(\bar{R}_j- \frac{1}{J}
    \sum_{j=1}^{J} \bar{R}_j\right)^2 
    \label{eq:between_cluster_social_variance}
\end{equation}

For comparison, the between-node $R$ variance in the original network, $\sigma^2_{\mathrm{R,bw-orig}}$, is computed using Eqn. \eqref{eq:between_cluster_social_variance} for $\mathcal{G}^{(1)}$, the input network to the overall network reduction methodology. 

The relative between-cluster $R$ variance is then defined as
\begin{equation}
    \Delta\sigma^2_{\mathrm{R,bw}} =
    \frac{
    \sigma^2_{\mathrm{R,bw}}-\sigma^2_{\mathrm{R,bw-orig}}
    }{
    \sigma^2_{\mathrm{R,bw-orig}}
    }
    \times 100\% 
    \label{eq:relative_between_social_variance}
\end{equation}
This metric measures the change in variance introduced by network reduction. A value close to zero indicates that the variance of $R$ in the reduced network is close to that of the original network. Negative values indicate that the variance has decreased: network reduction has ``smoothed'' the scarce resource scores across clusters, potentially resulting in a loss of decision-relevant information. Positive values indicate that the variance has increased: network reduction has ``amplified'' the difference in scarce resource scores across clusters, which may make it easier to identify/classify vulnerable populations.

\section{Experimental Results}
The reduction methods are evaluated on the Galveston G3 network. The original network has $1,457$ nodes. The Kron reduction method eliminates zero-injection nodes, resulting in an equivalent network with 600 nodes, a 59\% reduction in network size. Since the reduced network maintains the voltage characteristics and spatial distribution in scarcity scores, it is used as the baseline for the other methods. Table~\ref{tab:results} summarizes the performance results. The scarcity-weighted and scarcity-constrained methods are denoted as $R$-weighted and $R$-constrained respectively.
\begin{table}[t]
\centering
\caption{Performance comparison of network reduction methods.}
\label{tab:results}
\setlength{\tabcolsep}{2pt}
\renewcommand{\arraystretch}{0.9}

\begin{tabular}{lcccccc}
\toprule
Method 
& $n_{\mathrm{red}}$ 
& $\rho$ 
& $\Delta V$
& $\sigma^2_{\mathrm{R,in}}$
& $\sigma^2_{\mathrm{R,bw}}$
& $\Delta\sigma^2_{\mathrm{R,bw}}$
\\
& 
& (\%) 
& (\emph{m}p.u.) 
& 
& 
& (\%) 
\\
\midrule
Kron          & 600  & -- & 0.0      & -     & 0.87 & -     \\
Relaxed Kron-I    & 284  & 53 & 17 & -     & 0.92 & 5  \\
Relaxed Kron-II     & 163  & 73 & 24 & -     & 0.64 & -26 \\
Opti-KRON     & 18   & 97 & 0.94  & 0.45 & 0.19 & -79 \\
$R$-weighted     & 10   & 98 & 2.1  & 0.60 & 0.15 & -83 \\
$R$-constrained   & 82   & 86 & 3.8  & 0.28 & 1.17 & 34 \\
\bottomrule
\end{tabular}
\begin{tablenotes}
\fontfamily{phv}\fontsize{9}{10.8}\selectfont
\item Note: Opti-KRON follows \textcite{chevalier2022optikron}.  A dash ``-'' indicates that the corresponding metric is not applicable for the specific method.
\end{tablenotes}
\vspace{-0.8em}
\end{table}

\subsection{Network Reduction Rate}
The relaxed Kron methods achieve the lowest reduction rates, limited by the physical structure of the network and classification into primary, secondary, and branch layers. While further reductions are possible, additional insights are needed into the network to further specify sets $\mathcal{K}$ and $\mathcal{R}$. The optimization-based methods are not limited by specified node sets, which permits further network reduction. Both Opti-KRON and the $R$-weighted methods achieve significant network reduction of $\rho=97\%$ and $\rho=98\%$ respectively. The $R$-constrained method achieves a lower reduction rate of $\rho=86\%$, reflecting the additional requirement to preserve spatial distribution of scarcity metrics in the $R$-dissimilarity constraints.

\subsection{Voltage Error}
\begin{figure}[!t]
    \centering
    \includegraphics[width=\linewidth]{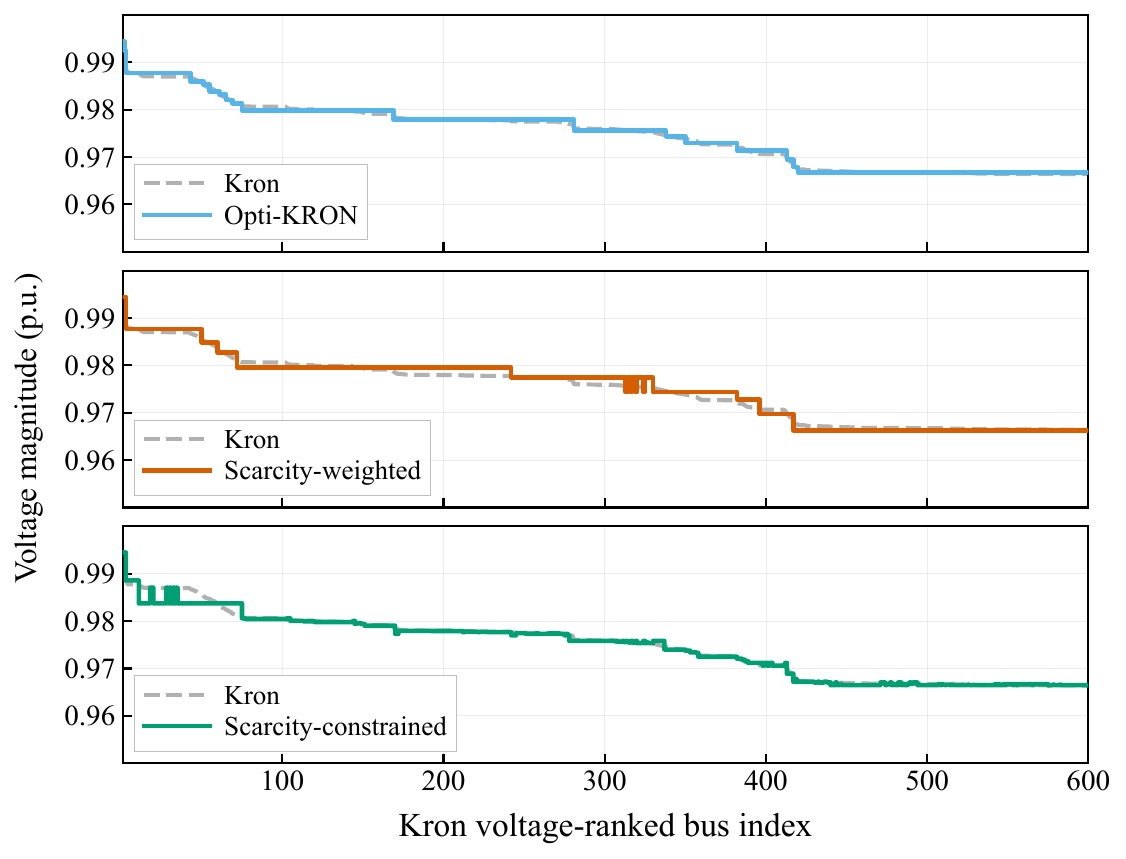}
    \caption{Voltage-profile comparison among the original and reduced networks.}
    \label{fig:faceted_voltage_profiles}
\end{figure}

The relaxed Kron reduction methods exhibit the highest voltage errors, with increasing error as additional layers of the network are eliminated. This is expected, as the method treats all eliminated nodes as zero-injection nodes and neglects any loads at these nodes. The optimization-based methods transfer loads from eliminated nodes to the representative node of the cluster and impose explicit voltage error constraints, resulting in significantly lower voltage errors. It should be noted that since the relaxed Kron methods do not cluster nodes, the voltage error metrics are computed differently. While the numerical values are not directly comparable, they provide a qualitative comparison across the method types.

Opti-KRON achieves the smallest voltage error, under 1\emph{m}p.u. The voltage errors are the highest for the $R$-constrained method, with an error of 3.8\emph{m}p.u. The $R$-weighted and $R$-constrained methods must balance multiple priorities and additional constraints imposed on the clustering of nodes. The tradeoff is clear: to preserve the spatial distribution of the scarcity scores, a nominal increase in voltage error is expected. Despite this tradeoff, the worst-case voltage error remains less than 0.01 p.u. across all nodes, illustrating that the proposed methods can be used on realistic distribution grids while still preserving critical voltage information. 

Figure~\ref{fig:faceted_voltage_profiles} plots the voltage profiles of the reduced networks against the Kron reduced network (which preserves the network voltages of the original G3 network). Nodes are ranked by decreasing voltage. All nodal voltages are above the voltage limit of 0.95 p.u. for the operating condition evaluated. The three optimization-based methods are able to preserve network voltages along the entire feeder. The Opti-KRON voltage profile remains close to the Kron reduced profile, with minimal deviation across the feeder. The $R$-weighted method exhibits slight deviations in the 0.97–0.98 p.u. range, where the reduced network overestimates the voltage at clustered nodes. For the $R$-constrained method, the voltage errors are most prominent near the substation.

\begin{figure*}[!t]
    \centering
    \subfloat[Opti-KRON]{%
        \includegraphics[width=0.325\textwidth]{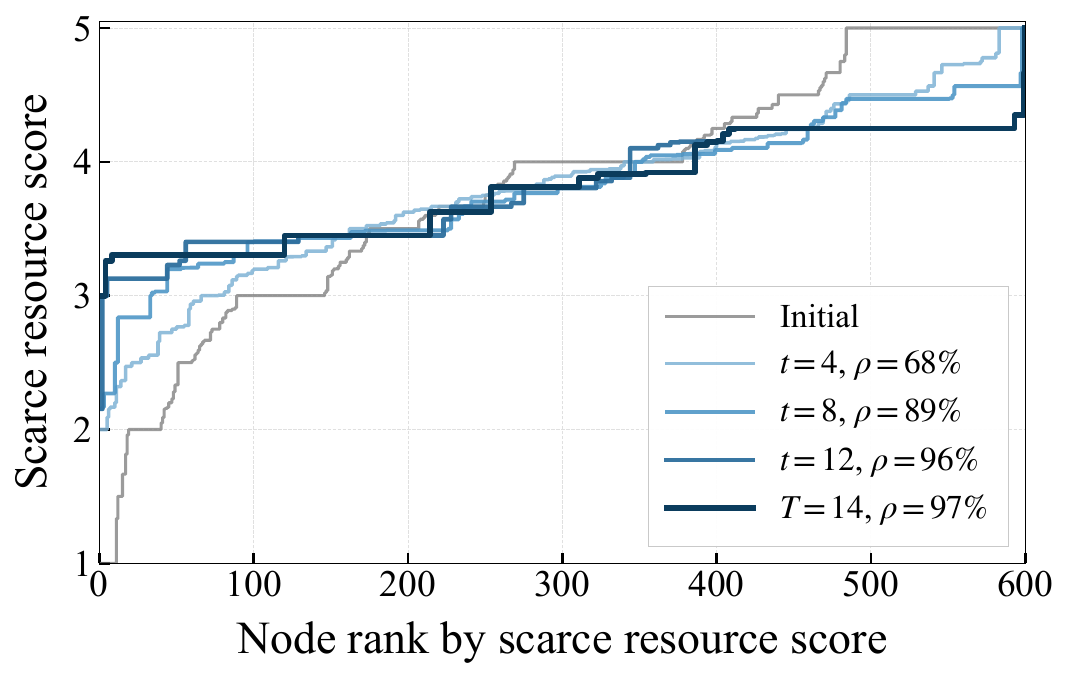}
        \label{fig:optikron_scarce_resource_duration_curve}
    }
    \hspace{-0.018\textwidth}
    \subfloat[$R$-weighted]{%
        \includegraphics[width=0.325\textwidth]{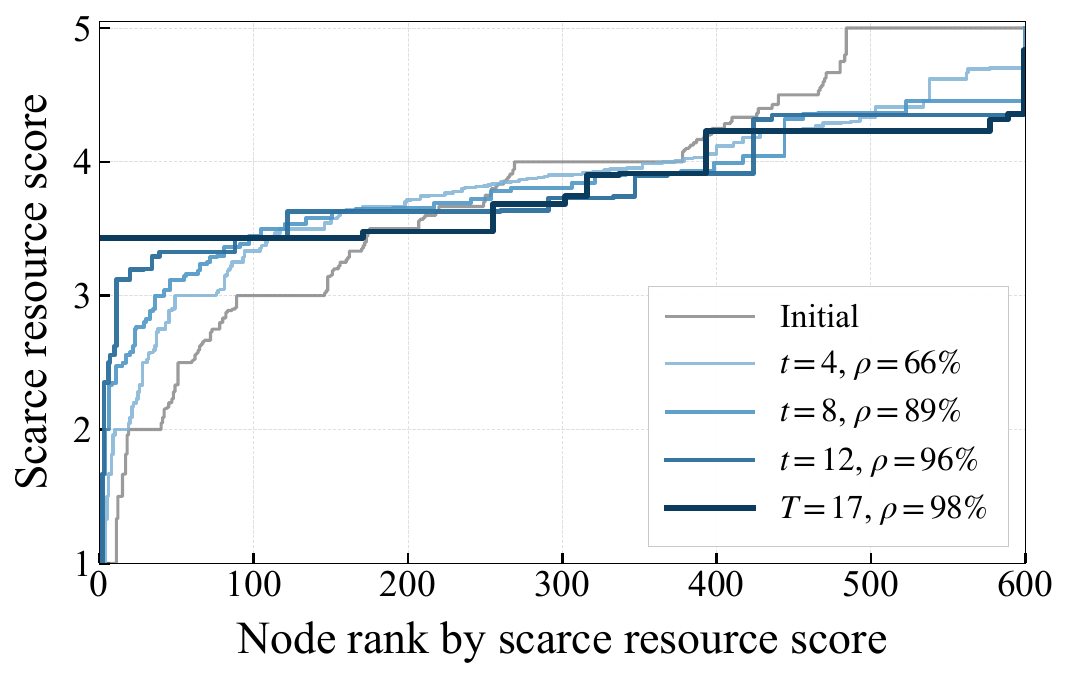}
        \label{fig:scarcity_weighted_scarce_resource_duration_curve}
    }
    \hspace{-0.018\textwidth}
    \subfloat[$R$-constrained]{%
        \includegraphics[width=0.325\textwidth]{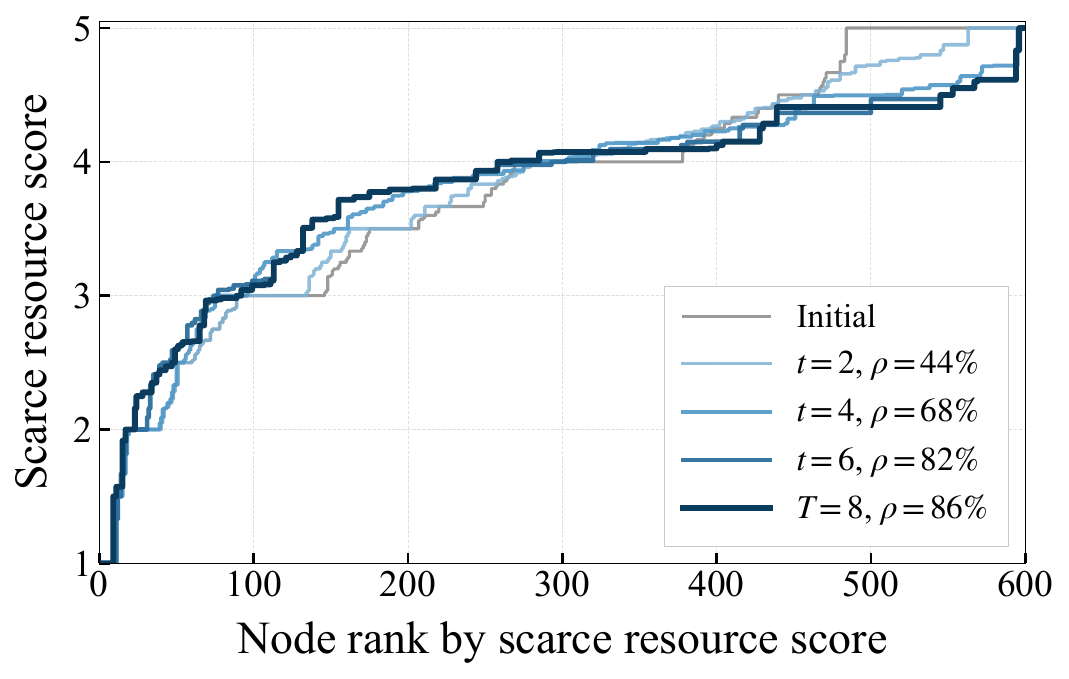}
        \label{fig:scarcity_constrained_scarce_resource_duration_curve}
    }
    \caption{Scarce resource score profiles for the optimization-based reduction methods, across optimization iterations, $t=\{1,\ldots,T\}$.}
    \label{fig:scarce_resource_duration_curves}
\end{figure*}

All three methods can preserve the voltages of the nodes near the end of the feeder, where the utility may be concerned with undervoltage conditions. However, the voltages near the substation and mid-line along the feeder have the highest voltage error, potentially resulting in challenges capturing the behaviour of voltage control devices such as on-load tap changers and voltage regulators in simulation. The voltage errors can be improved by modeling the line losses over the removed lines. Although the load is transferred from the eliminated node to the representative node, line parameters are not updated to reflect the removal of lines between kept and eliminated nodes. Additional simulations are needed to determine the sensitivity of voltage control devices, and to evaluate the voltage errors under different loading conditions, with distributed generation, and on additional feeders. These simulations should be used to further calibrate the user-defined parameters ($\alpha, \beta,\gamma,\phi,\omega$).

\subsection{Spatial Distribution of Scarce Resource Scores}

The within-cluster $R$ variance, $\sigma^2_{\mathrm{R,in}}$, is calculated for the three optimization-based methods. The $R$-weighted method has the highest variance, corresponding with the highest network reduction. The model incentivizes nodes with low $R$ scores to be kept, but does not impact how eliminated nodes are clustered. The Opti-KRON method outperforms $R$-weighted on all metrics: achieving lower voltage errors, lower $\sigma^2_{\mathrm{R,in}}$, and comparable $\sigma^2_{\mathrm{R,bw}}$. These results show that the weighted objective function is not sufficient to preserve the spatial distribution of $R$ scores and would induce similar averaging effects as Opti-KRON. In comparison, the $R$-constrained method achieves the lowest $\sigma^2_{\mathrm{R,in}}$, indicating that the average cluster-level $R$ scores are comparable to the other nodes in the cluster and may better represent the overall cluster. The between-cluster $R$ variance, $\sigma^2_{\mathrm{R,bw}}$ increases from the Kron network to the $R$-constrained network: through network reduction, the existing variance of $R$ scores is amplified, which may make it easier to identify/classify vulnerable populations.

Figure~\ref{fig:scarce_resource_duration_curves} plots the profile of scarce resource scores of the network across the iterations of network reduction. Nodes are ranked by increasing $R$ value for the Kron reduced network. The initial curve represents the $R$ profile of the input network, $\mathcal{G}_{Kron}$. As the iterations progress, the network reduction averages nodal $R$ scores, flattening the overall profile. Both the Opti-KRON and $R$-weighted methods exhibit this averaging behaviour in Fig.~\ref{fig:optikron_scarce_resource_duration_curve} and Fig.~\ref{fig:scarcity_weighted_scarce_resource_duration_curve} respectively. Vulnerable nodes with scarce resource scores below 3 are averaged upward to the 3-3.5 range, and nodes with high scarce resource scores above 4.5 are averaged downward to 4.2. These plots highlight an imbalance in information loss of the low resource and high resource nodes: while both are impacted by network reduction, the cluster-level representation for the low resource nodes experience higher information loss than the high resource nodes. The $R$-constrained method, in comparison, preserves the $R$ profile through the network reduction process. As shown in Fig.~\ref{fig:scarcity_constrained_scarce_resource_duration_curve}, the $R$-constrained method preserves the $R$ scores of vulnerable nodes with scarce resource scores below 3, while scores between 3 to 4 are aggregated upward to 4, and nodes with a score of 5 are aggregated downward to 4.5. 

The methods can also be compared at similar levels of network reduction, at iteration $t=8$ where $\rho\in[86,89]\%$. At this level, both Opti-KRON and the $R$-weighted methods have already impacted the $R$ profile, especially at low values of $R$, while the $R$-constrained method has preserved the $R$ profile. By incorporating $R$-dissimilarity constraints into the formulation, the $R$-constrained method better preserves spatial distribution of community impact metrics and can differentiate among clusters while limiting voltage errors and achieving significant network reduction.

\begin{figure}[!t]
    \centering
    \includegraphics[width=0.9\linewidth]{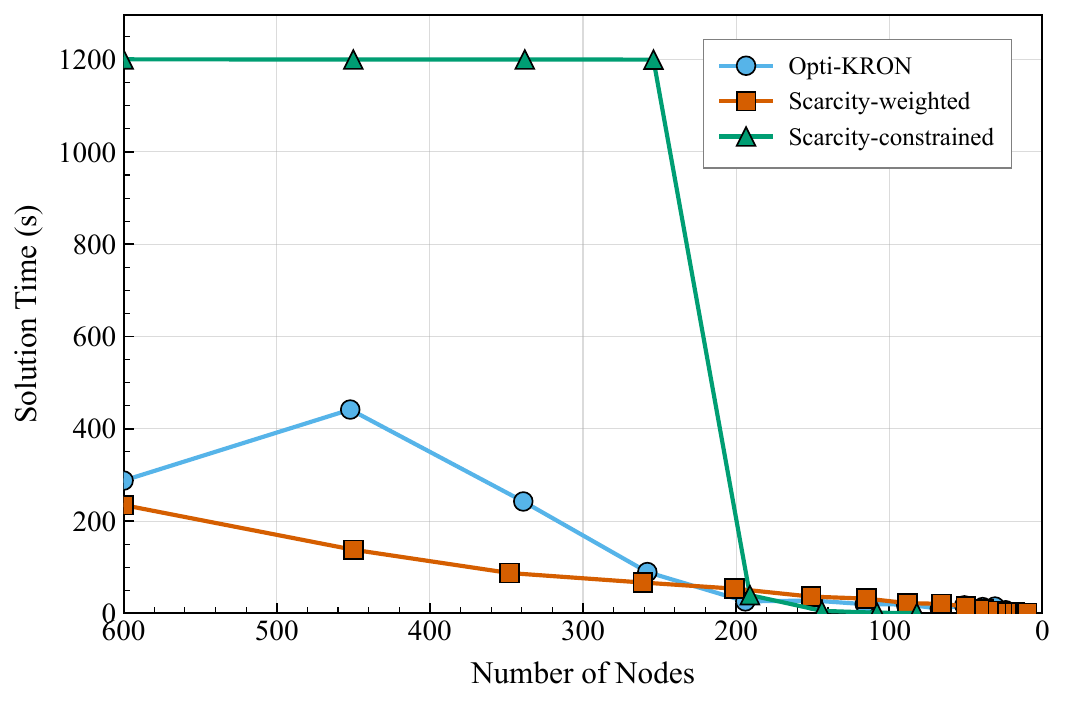}
    \caption{Solve time per iteration.}
    \label{fig:iteration_solution_time}
\end{figure}

\subsection{Computational Performance}

The model solve time reflects the computational complexity of each optimization-based reduction method, applied to the 600-node network. Figure~\ref{fig:iteration_solution_time} plots the computational time per iteration against the number of nodes of the original network, i.e., starting at 600 nodes and decreasing as the network reduction progresses. The curves generally show a decrease in solve time as the network size is progressively reduced. The Opti-KRON method takes a total of 33 minutes and 14 iterations. Early iterations are computationally expensive due to the large assignment matrix, $A$. Significant time is spent improving the optimality bounds and certifying solution quality to reduce the optimality gaps. The $R$-weighted method has the shortest total solve time of 12 minutes, taking 17 iterations. The addition of the $R$-weights provides a stronger incentive for network reduction when using the same $\alpha$ parameter value. This method quickly guides the solver toward higher-reward removals, shortening the solve time compared to Opti-KRON, even though the number of eliminated nodes per iterations is similar. The $R$-constrained method has the longest solve time, at 81 minutes and 8 iterations. The first few iterations require 20 minutes each, reaching the prescribed solver time limit. This is expected because of the additional constraints and objective function term on the $R$-dissimilarity. This highlights scalability as an area of future work, to extend the proposed framework to larger distribution networks. Despite this higher computational cost, the $R$-constrained method remains tractable for the 600-node Galveston network and achieves an 86\% network reduction while explicitly preserving community-impact characteristics.

\section{Conclusion}
This paper presents a community impact-centered network reduction method for power systems resilience planning. By incorporating community impact metrics in the objective function and introducing metric similarity constraints, the proposed method extends conventional network reduction approaches to jointly preserve the underlying power physics and the spatial distribution of community data present in the original network. We apply the method on a realistic distribution network model of Galveston Island, attaining an 86\% reduction rate with worst-case voltage magnitude errors under 0.01 p.u., while minimizing within-cluster variance and enhancing between-cluster variance of community impact metrics. The simulation results show our method is able to balance the competing objectives of network reduction and accurate representation of the underlying data. Future work will integrate the proposed reduced networks into decision optimization models to quantify the influence of network reduction on resilience-planning outcomes. 

\printbibliography
\end{document}